\documentclass[reprint,
 amsmath,amssymb,
 aps,
 prx
]{revtex4-2}

\usepackage{graphicx}
\usepackage{dcolumn}
\usepackage{bm}
\usepackage{physics}
\usepackage{hyperref}

\begin{document}

\preprint{Nature Communications}

\title{
\textbf{Harmonic origin of local symmetry breaking in PbTe revealed using machine-learned potentials}
}

\author{Seongjoo Jung}
\email{seongjoo.jung@mat.ethz.ch}
\author{Nicola A. Spaldin}%
\email{nicola.spaldin@mat.ethz.ch}

\affiliation{
Department of Materials, ETH Z\"{u}rich, Wolfgang-Pauli-Strasse 27, 8093 Z\"{u}rich, Switzerland
}

\begin{abstract}
We use machine-learned interatomic potentials combined with interatomic force constant models to show that finite-range polar correlations in PbTe are driven by long-range harmonic interactions rather than lattice anharmonicity. Our molecular dynamics simulations reveal a spatial polar correlation range exceeding 50 Å at low temperatures and decreasing with increasing temperature. Selective interatomic force constant mapping demonstrates that crystal anharmonicity usually associated with the formation of polar nanoregions suppresses polar correlations. Instead, long-range harmonic interactions extending to 13 Å are necessary to reproduce the spatial coherence of local dipoles. In reciprocal space, these extended harmonic interactions correspond to high curvature of the transverse optical (TO) phonon branch at the $\Gamma$ point. Comparison with isostructural NaCl connects these extended interactions to local underscreening of predominantly Coulombic forces. Our results provide a key microscopic description of finite-range polar correlations and a framework for understanding local symmetry breaking in centrosymmetric materials.
\end{abstract}

\maketitle

\section{Introduction}
The presence of an inversion center (centrosymmetry) in a crystal strictly prohibits macroscopic properties such as pyroelectricity, chirality, piezoelectricity, and even-order nonlinear optical responses \cite{halasyamani1998noncentrosymmetric}. Ferroelectricity is a prominent manifestation of macroscopic inversion symmetry breaking, characterized by a switchable spontaneous polarization usually emerging below a distinctive transition temperature. However, local structural fluctuations can break inversion symmetry over finite spatiotemporal scales even while the crystal preserves its average high-symmetry phase. Local symmetry breaking can produce optical, electronic, and electromechanical responses that are not apparent from the average crystal structure alone \cite{kusainova2001ferroelectric,manley2016giant,niesner2018structural}. 

Distortions associated with such spontaneous local symmetry breaking often occur in ferroelectrics above their ferroelectric transition temperature, and are associated with the soft (low frequency) transverse-optical (TO) phonons. They lead to phenomena such as polar nano- and mesodomains \cite{zhang2026imaging,fauque2022mesoscopic}, polarization density waves \cite{orenstein2025observation}, order--disorder ferroelectric transitions \cite{blinc2004order},
and \textit{emphanisis} \cite{bovzin2010entropically,knox2014local} and are known to be strongly correlated with crystal anharmonicity, including phonon--phonon, strain--phonon, and flexoelectric coupling \cite{marais1991phenomena,rubtsov2000crossover,guzman2023lamellar,orenstein2025observation}.

PbTe is an example that maintains a high-symmetry cubic phase at all temperatures, lacking a conventional phase transition to a macroscopically polar state; extrapolation of the softening TO phonon frequency would suggest a ferroelectric phase transition at negative temperature. Nevertheless, real-space pair distribution function (PDF) analyses reveal anomalous, non-Gaussian broadening of the peaks, alongside a continuous increase in the asymmetry of nearest-neighbor Pb-Te pair correlations upon heating \cite{bovzin2010entropically}.
Extending this local structural model to three dimensions, single-crystal diffuse scattering and three-dimensional difference pair distribution function mapping have demonstrated an unusual, step-like decay of displacive pair correlations along the $\langle 100 \rangle$ crystallographic directions \cite{sangiorgio2018correlated}.

The development of a non-Gaussian distribution of Pb--Te bond lengths, together with the characteristic splitting of the optical phonon peak \cite{jensen2012lattice}, has been interpreted as a direct consequence of lattice anharmonicity \cite{shiga2014origin} mediated by strong coupling between transverse optical and longitudinal acoustic (LA) phonons \cite{delaire2011giant} and the associated local symmetry breaking. In contrast, no Pb off-centering is observed using x-ray absorption fine structure spectroscopy \cite{keiber2013lead}. Also, first-principles-derived molecular dynamics study suggests that crystal anharmonicity itself does not necessarily lead to the formation of fluctuating local dipoles \cite{chen2014first}. These observations challenge a purely phonon--phonon scattering-based interpretation and suggest that additional microscopic mechanisms may contribute to the emergence of correlated local polar distortions.

\begin{figure*}
\includegraphics[width=1.00\textwidth]{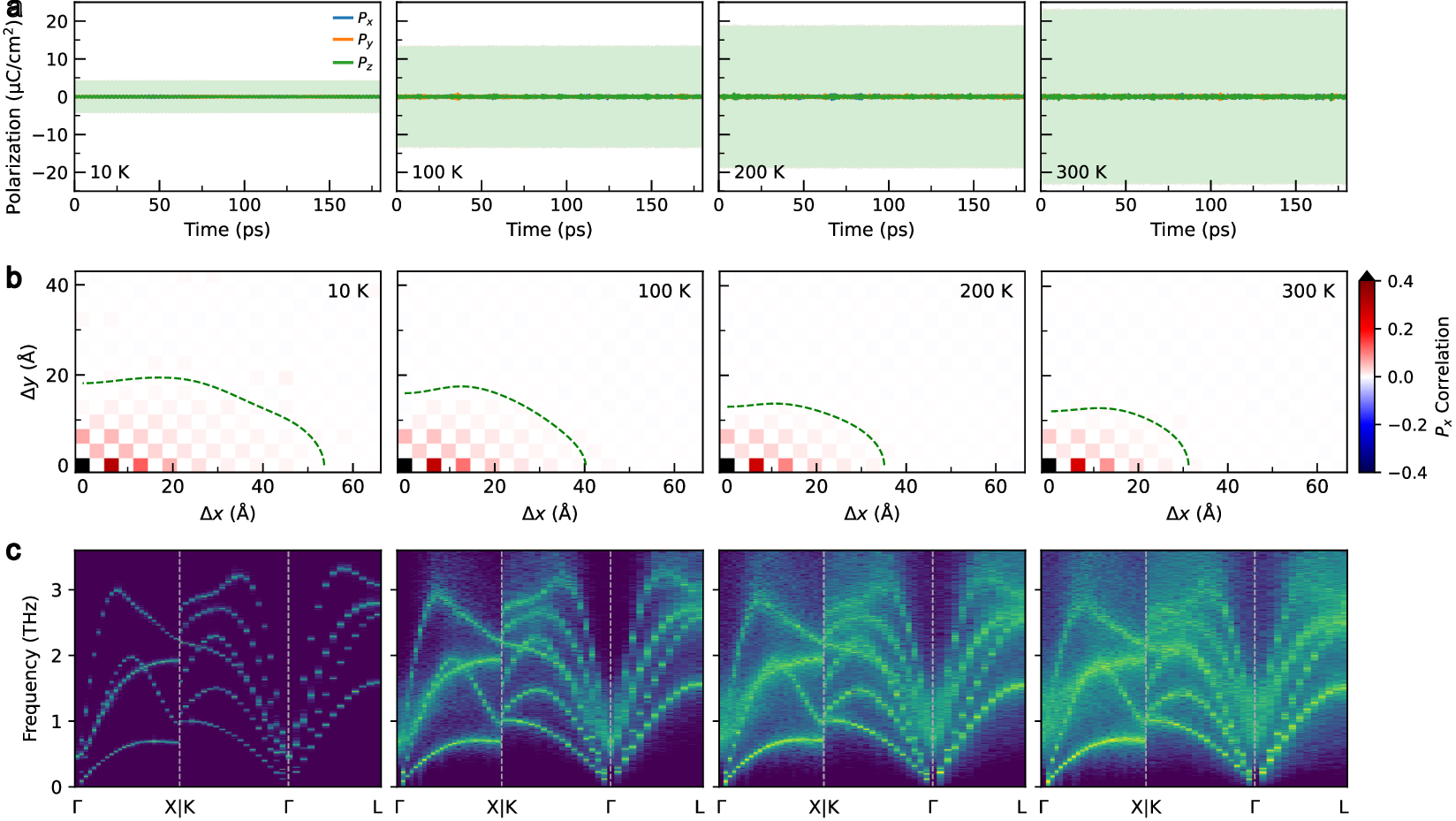}
\caption{Molecular dynamics simulations of PbTe using a machine-learned potential at temperatures from 10 to 300 K. (a) Average (solid lines) and standard deviation (shaded regions) of local polarization in the NpT ensemble at a pressure of $-15$ kbar. (b) The $P_xP_x$ component of the pair correlation matrix between neighboring octahedral centers in the $xy$ plane, showing the temperature-dependent range of spatial polar correlations. Each square in the heat map represents a local polarization center, corresponding to the position of a Pb atom. (c) Spectral energy density plots obtained from NVE simulations, illustrating anharmonic phonon scattering behavior with increasing temperature. While the higher-temperature trajectories exhibit TO mode hardening, peak broadening, and splitting, the lower-temperature trajectories, where polar correlations are strongest, show no clear signs of strong anharmonicity.}
\label{fig:MLIP}
\end{figure*}

In this work, we aim to reconcile these conflicting observations by identifying the microscopic origin of correlated local dipoles and establishing materials-design principles governing their emergence. We begin with the development of  machine-learned interatomic potentials (MLIPs) trained on density functional theory (DFT) data \cite{schutt2017schnet,gasteiger2020directional}, which enable molecular dynamics (MD) simulations at unprecedented length and time scales providing access to local structural correlations with high resolution. MD simulations based on MLIPs offer limited physical insight into the underlying interactions, so we construct an interatomic force constant (IFC) model that reproduces the MD behavior of the MLIP and provides a physically interpretable description of the underlying interatomic interactions. This combination of large-scale MLIP simulations and an interpretable force-constant framework allows us to directly connect the observed local structural correlations to their microscopic origin and identify the key factors governing the emergence of correlated local dipoles.

We find that finite-range polar correlations in PbTe are driven not by thermal lattice anharmonicity, but by long-range harmonic interatomic interactions \cite{lee2014resonant}. Surprisingly, higher-order crystal anharmonicity instead acts to suppress spatial dipole alignment.
In reciprocal space, this long-range harmonic force profile translates to high curvature of the TO phonon branch around the $\Gamma$ point. By comparing the DFT-derived force constants and differences in electric potential between PbTe and NaCl, we demonstrate that the range of second-order harmonic IFCs from local electrostatic underscreening is a fundamental parameter governing the emergence and spatial scale of local polar order in non-polar phases.

\section{Polar correlation from MLIP trajectories}

To explore long-range correlations of local polarization in PbTe at various temperatures, we performed molecular dynamics simulations of a $24\times24\times24$ supercell of conventional PbTe unit cells (110,592 atoms) using a custom MACE potential trained on DFT data \cite{batatia2022mace,Batatia2022Design}. The DFT calculations employed the local density approximation (LDA), including spin--orbit coupling and an applied negative pressure. We found that this approach provided the best agreement with the target experimental low-temperature properties, including a lattice constant of 6.430 Å (0 K) \cite{minikayev2021thermostructural}, a band gap of 0.19 eV (30 K) \cite{strehlow1973compilation,gibbs2013temperature}, a transverse optical (TO) phonon frequency of 0.5 THz (4.2 K) \cite{tacke1982dielectric,kawahala2023thickness}, an ionic contribution to the dielectric constant of approximately 1300 (4.2 K) \cite{tacke1982dielectric}, and Born effective charges of $\pm 6.5\,e$ (300 K) \cite{cochran1966crystal,ravindra1980properties} (see Supplementary Information).

Fig.~\ref{fig:MLIP} shows the properties derived from the resulting trajectories at temperatures from 10 to 300 K. Local polarization was calculated for each Pb-centered octahedron as the dipole moment formed between the cation and anion centers, using their corresponding Born effective charges, divided by the octahedron volume. Panel (a) shows the average (solid lines) and standard deviation (shaded regions) of local polarization over the entire trajectory at each of four temperatures in the NpT ensemble at a pressure of $-15$ kbar. Consistent with previous studies, no net polarization emerges, and the magnitudes of local polarization increase with increasing temperature, as reflected in the standard deviations.

Panel (b) shows the $P_xP_x$ component of the pair correlation matrix between neighboring octahedral centers in the $xy$ (001) plane. Each colored square represents a Pb-centered octahedron to which a local polarization is assigned. Even at 300 K, correlations in the direction parallel to the displacement persist up to $\sim$30 Å, whereas perpendicular correlations persist up to $\sim$10 Å, consistent with a previous molecular dynamics study \cite{sangiorgio2018correlated}. However, in contrast to previous reports of correlated local dipoles emerging at high temperatures, the range of polar correlations monotonically increases with decreasing temperature, reaching over 50 Å at 10 K.

The appearance of nanoscale to mesoscale polarization in quantum paraelectrics such as SrTiO$_3$ is often attributed to coupling between optical and acoustic modes, or to flexoelectric interactions \cite{fauque2022mesoscopic,orenstein2025observation}. PbTe is known to exhibit similar anharmonic coupling between TO and LA phonons \cite{delaire2011giant}. To search for anharmonic phonon scattering behavior, we obtained spectral energy density (SED) plots from molecular dynamics simulations in the NVE ensemble; the results are shown in Fig.~\ref{fig:MLIP} panel (c). Similar correlation behavior was observed in these trajectories as in the NpT trajectories, indicating no significant contribution from phonon--strain coupling.

\begin{figure}
\includegraphics[width=0.49\textwidth]{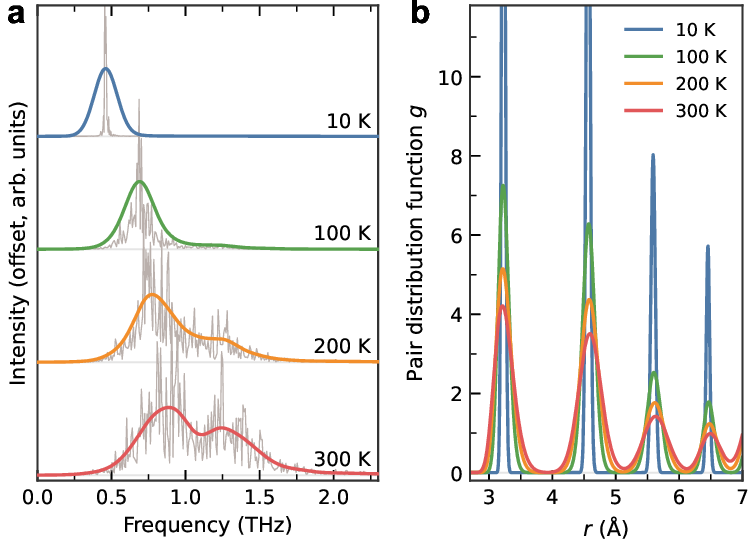}
\caption{(a) Calculated temperature-dependent spectral energy density profiles illustrating the $\Gamma$-point TO mode frequency distribution. (b) Pair distribution function obtained from MLIP-based MD trajectories at temperatures between 10 K and 300 K. }
\label{fig:TO}
\end{figure}

As expected, strong anharmonic phonon behavior is observed in the high-temperature trajectories. Fig.~\ref{fig:TO}(a) shows the $\Gamma$-point TO mode frequency distribution obtained from the SED plots. As temperature increases, TO mode hardening, phonon peak broadening due to increased scattering, and peak splitting are observed.
However, clear signatures of strong TO--LA coupling, such as avoided crossings and corresponding dips in the acoustic mode, are absent. In particular, at lower temperatures where polar correlations are most prominent, no signs of anharmonicity are observed.

We further analyze the PDFs obtained from the MLIP trajectories. Fig.~\ref{fig:TO}(b) presents the PDF up to 7 Å from MLIP-based MD trajectories at temperatures ranging from 10 K to 300 K, averaged over all atoms and time steps in each trajectory. The calculated PDFs reproduce the experimentally observed peak positions and their overall temperature-dependent evolution reported previously \cite{bovzin2010entropically}. In particular, the nearest-neighbor peak progressively broadens and develops an asymmetric line shape with increasing temperature.

\begin{figure*}
    \includegraphics[width=1.00\textwidth]{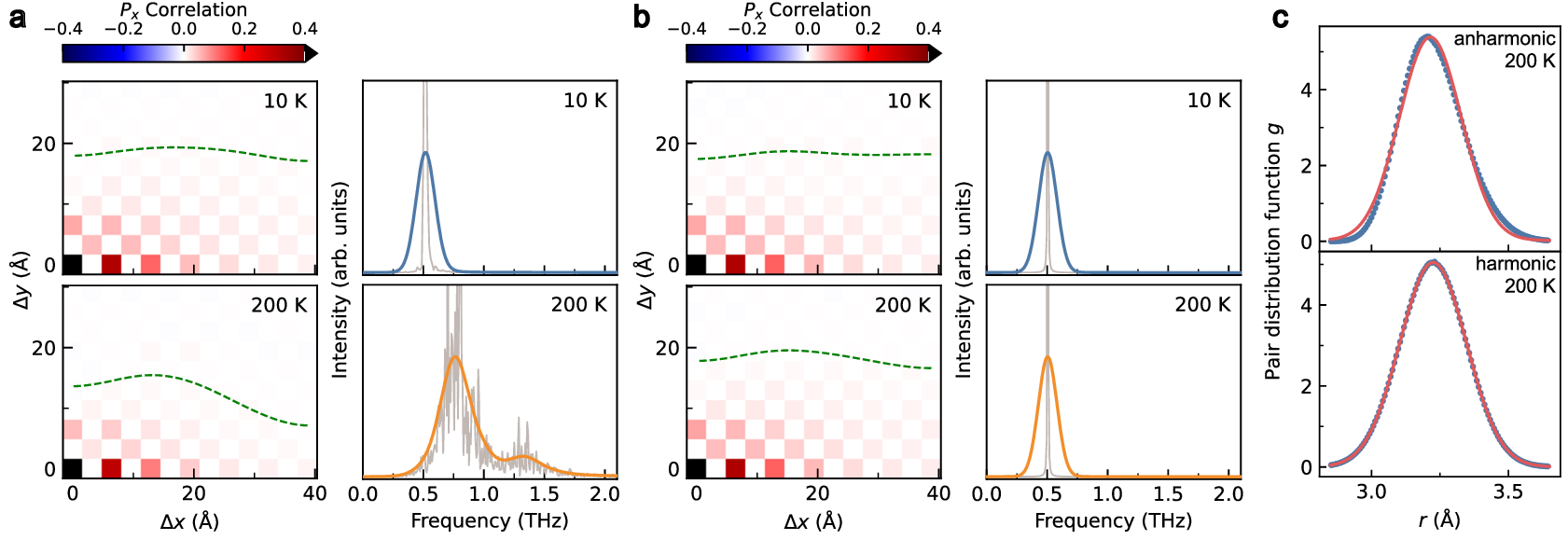}
    \caption{Effect of lattice anharmonicity on spatial polar correlations and phonon dynamics calculated using mapped interatomic force constants. (a) Spatial polarization correlation maps ($P_x$ component) and $\Gamma$-point SED profiles at 10 K and 200 K, including higher-order anharmonic interactions. (b) Spatial polarization correlation maps and SED profiles restricted to purely harmonic interactions. The purely harmonic model yields infinitely sharp, temperature-invariant TO phonon peaks. (c) Nearest-neighbor PDF peaks and corresponding Gaussian fits obtained from MD trajectories generated using anharmonic and harmonic IFC potentials.}
    \label{fig:IFC}
\end{figure*}

\section{Persistence of spatial coherence in the harmonic limit}

While MLIPs enable accurate molecular dynamics studies at unprecedented scales, they offer limited insight into the underlying mechanisms. To identify the origin of the observed behavior, we map the MLIPs onto IFCs $\Phi$, which are the coefficients in the expansion of the crystal energy $E$ in terms of ionic displacements $u$:
\begin{equation}
    E = E_0 + \Phi_i^\alpha u_i^\alpha + \frac{1}{2!}\Phi_{ij}^{\alpha\beta} u_i^{\alpha}u_j^{\beta} + \frac{1}{3!}\Phi_{ijk}^{\alpha\beta\gamma} u_i^{\alpha}u_j^{\beta}u_k^{\gamma} + \dots 
\end{equation}
Here, $\alpha, \beta, \gamma$ denote Cartesian indices, and $i, j, k$ denote atomic indices within the supercell using the Einstein summation notation. We compare the molecular dynamics trajectories generated using the MLIP with those generated using the corresponding IFCs for a $12\times12\times12$ supercell of 8-atom conventional PbTe unit cells (13,824 atoms). We find that the local correlation behavior in the MLIP trajectories is correctly recovered only when harmonic interactions extending to the 14th nearest neighbor (12.9 Å in the symmetric configuration without displacements) are included. This range spans two conventional unit cells and corresponds to 78 degrees of freedom. For anharmonic interactions, we include third-order IFCs extending to the second nearest neighbor (36 degrees of freedom) and fourth-order IFCs extending to the first nearest neighbor (4 degrees of freedom).

Fig.~\ref{fig:IFC}(a) shows the spatial polar correlations and TO mode frequencies obtained from the IFC trajectories and SED respectively, at 10 K and 200 K. Although the limited simulation cell size restricts the range over which the spatial correlations can be accurately captured, the simulations reproduce the expected temperature-dependent trends. In particular, increasing the temperature reduces the spatial extent of the polar correlations, while the TO mode peak exhibits broadening and splitting.

Panel (b) presents the corresponding results from molecular dynamics simulations in which anharmonic contributions from the third- and fourth-order IFCs are excluded. The resulting TO mode spectra exhibit the characteristic behavior of a purely harmonic potential energy surface. As expected, phonon lifetimes become effectively infinite, resulting in sharp peaks in the SED at both temperatures. Furthermore, neither TO mode hardening nor peak splitting is observed with increasing temperature. In practice, the intensity-normalized peaks are identical at both temperatures for the harmonic potential.

Notably, the polar correlation maps and sharp TO peaks in panel (b) show that nanoscale spatial polar correlations persist even when all anharmonic contributions from the third- and fourth-order IFCs are excluded. There is no qualitative difference between the correlation behavior at 10 K and 200 K, and the weak temperature dependence of the correlation profiles further suggests that changes in the thermal displacement amplitudes do not, by themselves, determine the spatial extent of the polar correlations. 
Comparing the 200 K data with and without anharmonic contributions in panels (a) and (b) reveals that crystal anharmonicity in fact {\it hinders} the formation of nanoscale polar correlations. This finding challenges the notion that anharmonicity is necessary to establish such correlations and instead identifies it as a mechanism that weakens correlations already supported by the underlying harmonic interactions.

\begin{figure*}
    \includegraphics[width=0.95\textwidth]{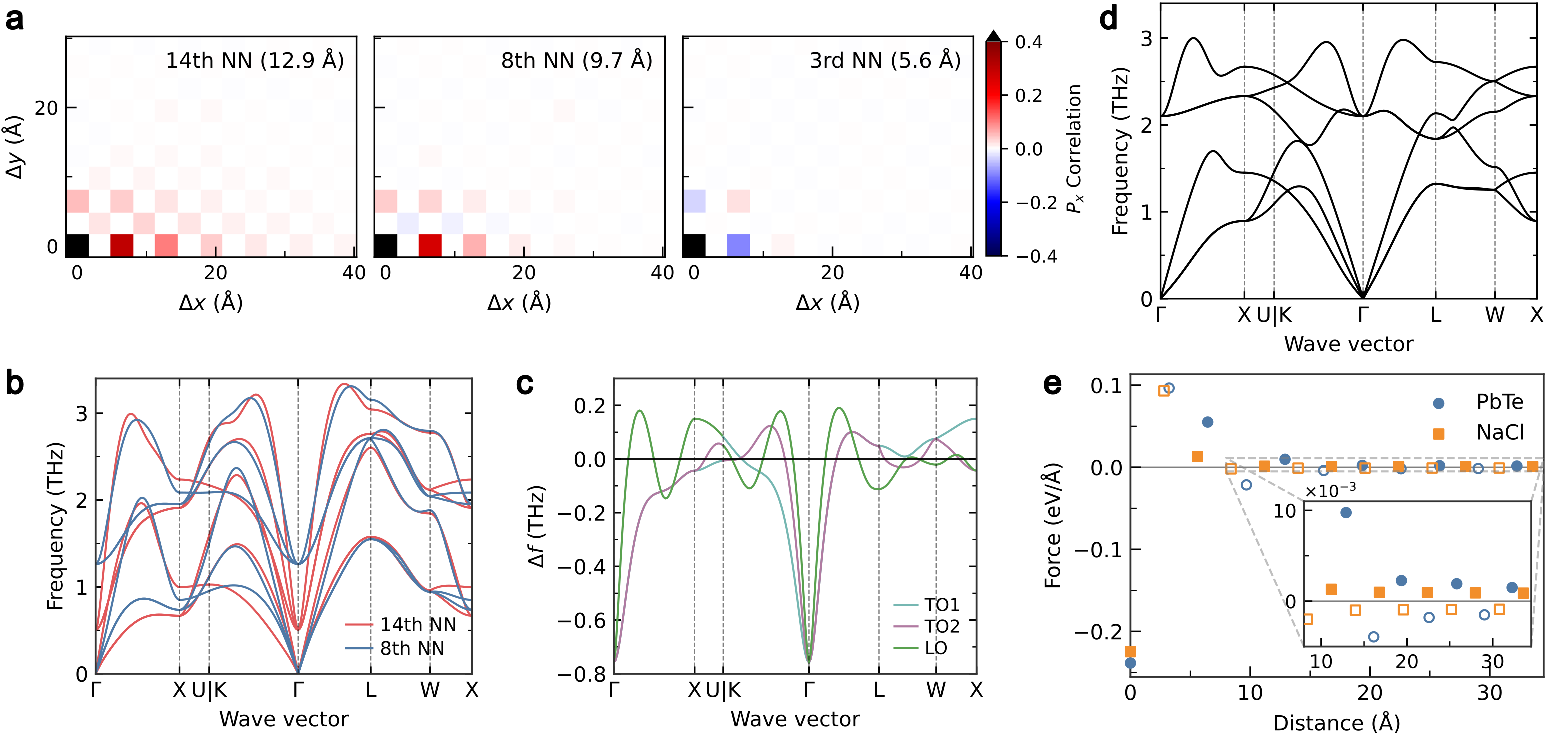}
    \caption{Role of long-range harmonic interactions in local polar correlations in PbTe. (a) Spatial pair correlation maps of the $P_x$ polarization component evaluated with harmonic interatomic force constant cutoffs restricted to the 14th (12.9 Å), eighth (9.7 Å), and third (5.6 Å) nearest neighbors. Truncating the interaction range systematically reduces the coherence of the polar nanoregions. (b) Harmonic phonon dispersion relations constructed using interaction cutoffs at the 14th and eighth nearest neighbors. (c) Frequency differences ($\Delta f$) between the two cutoff models. (d) Phonon dispersion of NaCl. (e) Real-space atomic forces along a cation--anion chain in response to a cation displacement of 0.1 Å. PbTe maintains substantial long-range force constants extending to nearly 20 Å, whereas in NaCl the harmonic restoring forces decay rapidly beyond the fourth nearest neighbor.}
    \label{fig:long_range}
\end{figure*}

Fig.~\ref{fig:IFC}(c) shows the nearest-neighbor PDF peaks obtained from IFC-based MD trajectories at 200 K using anharmonic and harmonic force constants, together with the corresponding Gaussian fits. The purely harmonic potential produces a degree of peak broadening comparable to that obtained with the anharmonic potential. However, the peak remains nearly perfectly Gaussian. In contrast, the PDF obtained with the anharmonic potential exhibits the characteristic asymmetry.

To summarize, lattice anharmonicity in PbTe induces TO peak hardening, broadening, and splitting, as well as PDF peak asymmetry. However, in contrast to previous narratives, anharmonicity suppresses local polarization correlations relative to the purely harmonic potential. 
The bond asymmetry originates from anharmonicity within a single potential well and does not require off-centering associated with a double-well-like potential.
Our results indicate that the asymmetric peak observed in the PDFs does not necessarily imply correlated fluctuations of the local polarization or the formation of polar nanoregions. Rather, anomalous bond length behavior and polar correlations are separate phenomena with different origins \cite{bridges2023local}.

\section{Spatial coherence from long-range harmonic interactions}

To further investigate the origin of nanoscale polar correlations, we repeat the molecular dynamics simulations at 200 K, retaining the anharmonic contributions but using shorter cutoffs for the second-order harmonic IFCs. Fig.~\ref{fig:long_range}(a) shows results for three maximum harmonic interaction ranges: the 14th nearest neighbor (12.9 Å), the eighth nearest neighbor (9.7 Å), and the third nearest neighbor (5.6 Å). We find that local polarization correlations disappear as the range of harmonic interactions decreases. With interactions extending to the eighth nearest neighbor, the directional preference for correlations parallel to the displacement is retained, but the range is much more limited.  In the $y$ direction, the weak off-diagonal interactions even allow for a weak anticorrelation, with anti-aligned local dipoles, so that a nanoregion with net correlated polarization does not form. With interactions restricted to the third nearest neighbor, there are even more limited correlations and anticorrelations, visible only within one unit cell range.

The second-order IFCs contain all the information about harmonic interactions in real space. In reciprocal space, the phonon frequencies and eigenvectors throughout the Brillouin zone contain the same information and can be obtained by Fourier transformation of the real-space IFCs \cite{born1954dynamical,gonze1997dynamical}. Figures~\ref{fig:long_range}(b) and \ref{fig:long_range}(c) show how the harmonic phonon frequencies respond to reduced IFC cutoff distances. Panel (b) shows the phonon dispersions for IFC cutoffs at the 14th and eighth nearest neighbors, and panel (c) shows the frequency differences between the two cases for the branches originating from the optical modes at the $\Gamma$ point. The phonon frequencies are similar throughout the Brillouin zone for the optical modes (differing by less than 0.2 THz), except at the $\Gamma$ point, where the frequencies shift by nearly $-0.8$ THz when harmonic interactions between the ninth and 14th nearest neighbors are included \cite{ju2018revisiting}.

The stiffness of the TO mode alone does not determine the spatial correlation range. Although reducing the range of harmonic interactions leads to both TO mode hardening and diminished spatial correlations, the loss of correlation arises from changes in the phonon dispersion rather than from the increase in the TO mode frequency itself. To demonstrate this, we scaled the PbTe IFCs up to the third nearest neighbor by a factor of 0.05 to reproduce the TO mode frequency of 0.5 THz obtained with IFCs extending to the 14th nearest neighbor. The resulting molecular dynamics trajectories exhibit no discernible spatial correlations in local polarization, despite the matched TO mode frequency (see Supplementary Information).

These results indicate that the emergence of correlated polar nanoregions is primarily driven by substantial long-range harmonic interactions in real space, rather than by crystal anharmonicity associated with third- and fourth-order IFCs. In reciprocal space, this behavior manifests as a pronounced curvature of the TO phonon branch near the $\Gamma$ point, characterized by a low frequency at $\Gamma$ and rapid hardening with increasing wavevector. As the range of harmonic interactions is reduced, the TO phonon branch becomes progressively flatter, accompanied by a loss of spatial correlations in local polarization.

\section{Electrostatic origin of the extended interactions}

Finally, we compare the DFT force constants of isostructural NaCl and PbTe in greater detail and provide an electrostatic interpretation of the long-range interactions in PbTe. The long-range interactions in PbTe have been proposed to originate from unique chemical bonding that involves both covalent directionality and features of electron delocalization \cite{waghmare2003first,bersuker2013pseudo,lee2014resonant,wuttig2018incipient,hempelmann2021long}. Fig.~\ref{fig:long_range}(d) shows our calculated phonon dispersion of NaCl, which exhibits a substantially flatter transverse optical (TO) phonon branch than that of PbTe. Notably, the phonon dispersion of PbTe obtained by truncating harmonic interactions beyond the third nearest neighbor closely resembles that of NaCl, highlighting the distinct spatial ranges of harmonic interactions in the two materials (see Supplementary Information). Fig.~\ref{fig:long_range}(e) shows the induced forces along the $+x$ direction in response to a displacement of a single cation by 0.1 Å along the same direction in PbTe and NaCl supercells. Supercell dimensions of $64.5 \times 12.9 \times 12.9$ Å and $67.2 \times 11.2 \times 11.2$ Å were used for PbTe and NaCl, respectively, corresponding to $10 \times 2 \times 2$ and $12 \times 2 \times 2$ supercells.

The blue circles (orange squares) indicate the forces in PbTe (NaCl) at a distance $d$ from the displaced cation along the cation--anion chain in the $x$ direction. Filled markers indicate cations, and unfilled markers indicate anions. The force can be written in terms of the IFCs as
\begin{equation}
    F_i^x = -\pdv{E}{u_i^x} = -\Phi_{0i}^{xx} u_0^{x} - \frac{1}{2!}\Phi_{00i}^{xxx} u_0^{x}u_0^{x} + \dots 
\end{equation}
where $u_0^{x}$ represents the displacement of the cation and its periodic images of the supercell. The small displacement and large supercell size provide insight into many converged second-order IFCs from a single force calculation. (See Appendix B for analysis of electric potential under periodic boundary conditions)

The forces in panel (e) illustrate the classical interpretation of the competition between short-range electron overlap repulsion and long-range Coulomb interactions \cite{cochran1959crystal,anderson1960qualitative}. For a positive cation displacement, the long-range Coulomb interaction exerts a force in the positive direction on cations and in the negative direction on anions, contributing to the softening of the TO phonon mode. In contrast, the restoring force on the displaced cation and the force due to electron overlap repulsion on its nearest neighbor oppose the long-range Coulomb forces, contributing to phonon hardening.

Interestingly, the cation self-interaction force constants and nearest-neighbor IFCs are similar for PbTe and NaCl. The differences become apparent starting from the next cation along the chain, corresponding to the fourth-nearest neighbor in the rocksalt structure. For NaCl, the forces become negligible from the fourth-nearest neighbor onward, whereas substantially larger forces persist over longer distances in PbTe. At the 14th-nearest-neighbor distance (12.9~\text{\AA} for PbTe and 11.2~\text{\AA} for NaCl), the IFC magnitude in PbTe is nearly an order of magnitude larger than that in NaCl. The long-range interaction in PbTe persists over distances approaching 20~\text{\AA}.

Although this long-range interaction is predominantly Coulombic, its magnitude cannot be explained solely by the large Born effective charges in PbTe ($\pm 7.1~e$) compared with those in NaCl ($\pm 1.1~e$). The strength of the Coulombic interaction depends not only on the product of the Born effective charges of the interacting ion pair but also inversely on the high-frequency (clamped-ion) permittivity. The calculated bulk high-frequency permittivity of PbTe (83.2 $\epsilon_0$) is also much larger than that of NaCl (2.6 $\epsilon_0$). 

\begin{figure}
    \includegraphics[width=0.45\textwidth]{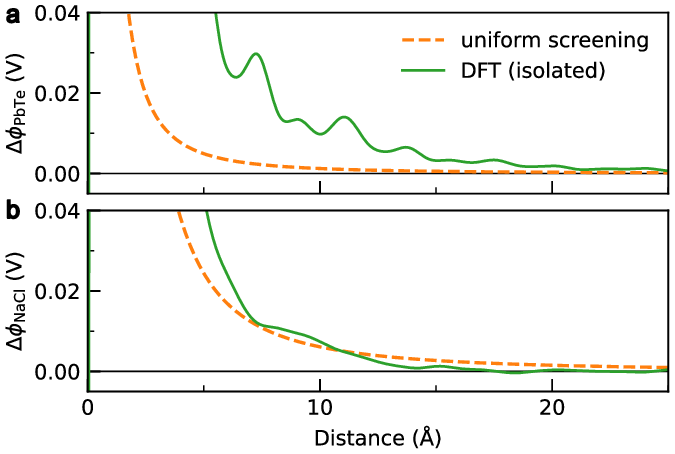}
    \caption{Electrostatic response to a local cation displacement. Displacement-induced potential changes in (a) PbTe and (b) NaCl are plotted as a function of distance along the $x$-axis for a cation displacement of 0.1 Å at the origin.}
    \label{fig:potential}
\end{figure}

Fig.~\ref{fig:potential} compares the DFT-calculated potential difference induced by cation displacement along the $x$-axis  \cite{castro2003solution} with the theoretical electrostatic potential of a dipole under the assumption of uniform dielectric screening for both PbTe and NaCl. For NaCl, the two profiles exhibit good overall agreement, whereas PbTe shows substantial deviations from the uniform-screening model up to $\sim$20 Å. Our analysis suggests that the previously identified bonding mechanisms primarily give rise to local underscreening of the electric field, while long-range interactions remain predominantly electrostatic in nature, with enhanced forces acting on both cations and anions.
 
\section{Conclusion}

We have identified the microscopic origin of finite-range correlated local polarization in cubic PbTe by combining large-scale machine-learning molecular dynamics with an interpretable interatomic force-constant description. Our simulations reveal pronounced directional polar correlations extending over several nanometers, with the correlation range increasing substantially upon cooling. Although PbTe exhibits strongly anharmonic phonon behavior at elevated temperatures, systematic removal of the third- and fourth-order force constants demonstrates that anharmonicity is not required for the formation of the nanoscale polar correlations. Instead, the correlations persist in the harmonic limit and are caused by the large spatial extent of the second-order interatomic force constants.

We further show that truncating the harmonic interactions progressively suppresses the local polar correlations. The corresponding changes in the phonon dispersion establish that the real-space range of the harmonic interaction is directly reflected in the curvature of the low-frequency TO branch near the $\Gamma$ point. 
Comparison with NaCl, which shares the same rocksalt structure, further demonstrates that the long-range harmonic interactions in PbTe, extending to nearly 20 Å, remain predominantly Coulombic and result from local underscreening.

These results indicate that the presence of a stable zone-center TO mode does not preclude substantial finite-range polar correlations. Moreover, such correlations do not necessarily require crystal anharmonicity or additional local minima, as commonly invoked in order--disorder transition models \cite{onodera1970dynamic}. We consider the correlation range captured in this study to be a lower bound. Although current MLIPs enabled large-scale calculations, the maximum interaction range is limited by factors such as the size of the training cell and the cutoff range, even with sufficient number of message-passing layers \cite{kosmala2023ewald}. We hope our study motivates further experimental and theoretical efforts to resolve the low-temperature local symmetry breaking and related material responses, as well as developments of MLIPs capable of accurately capturing long-range harmonic interactions.

\begin{acknowledgments}
We thank Dr. Hiroki Ueda for discussions and comments. We acknowledge the Euler cluster at ETH Zurich and the Swiss National Supercomputing Center (CSCS) project No. lp61 for providing resources that contributed to this research. This work was supported by the Swiss National Science Foundation under Grant No. 225790.
\end{acknowledgments}

\appendix

\section{Methods}
\subsection{DFT}

For PbTe, noncollinear spin-polarized periodic density functional theory (DFT) calculations were performed using the Vienna Ab Initio Simulation Package (VASP) 6.4.3 \cite{kresse1996efficient}, including spin--orbit coupling. Exchange--correlation effects were treated within the local density approximation (LDA) \cite{perdew1981self}. Interactions between ionic cores and valence electrons were described using the projector augmented-wave (PAW) method \cite{blochl1994projector,kresse1999ultrasoft}. A plane-wave kinetic energy cutoff of 450 eV was employed for the Kohn--Sham orbitals. The following valence electron configurations were used: Pb ($5d^{10}6s^26p^2$) and Te ($5s^25p^4$). The convergence threshold for residual ionic forces in structural relaxations was set to 0.001 eV/Å. For NaCl, both spin polarization and spin--orbit coupling were excluded from the DFT calculations. Exchange--correlation effects were treated using the PBEsol functional \cite{perdew2008restoring} within the generalized gradient approximation (GGA). A plane-wave kinetic energy cutoff of 350 eV was employed, with valence electron configurations of Na ($2p^63s^1$) and Cl ($3s^23p^5$).

\subsection{MLIP training}
On-the-fly training of a machine-learned interatomic potential (MLIP) \cite{jinnouchi2019phase} was conducted via ab initio molecular dynamics (AIMD) simulations using VASP. The AIMD calculations were performed for a 64-ion PbTe supercell (a $2\times2\times2$ supercell of conventional unit cells) in the isothermal--isobaric (NpT) ensemble using a Langevin thermostat for both atomic and lattice degrees of freedom, with target temperatures between 120 and 800 K, an applied pressure of $-11$ kbar, and a time step of 2.0 fs. A $6\times6\times6$ Monkhorst--Pack $k$-point mesh \cite{monkhorst1976special} was used to sample the Brillouin zone. Self-consistent electronic minimization was performed with an energy convergence threshold of $1.0\times10^{-8}$ eV. Partial electronic occupancies were treated using Gaussian smearing with a broadening width of 0.01 eV.

The PbTe interatomic potential was trained using the ScaleShiftMACE architecture within the MACE framework \cite{batatia2022mace,Batatia2022Design}. Two interaction layers, a correlation order of 3, a maximum angular momentum of $\ell_{\max}=2$, and a radial cutoff of $6.0~\text{\AA}$ were used. The model was trained for up to 3000 epochs with a batch size of 64, using energy, force, and stress loss weights of 5.0, 50.0, and 1.0, respectively. The AMSGrad optimizer was employed with a weight decay of $5\times10^{-4}$, while early stopping was applied with a patience of 300 epochs. An exponential moving average (EMA) with a decay factor of 0.99 and stochastic weight averaging (SWA), initiated at epoch 2500, were used to improve model stability. Training was performed on a CUDA-enabled GPU using single-precision arithmetic. Separate training, validation, and test datasets were used to evaluate model performance. The trained model achieved a root-mean-square error (RMSE) of 0.9 meV/atom for energy, a force RMSE of 17.6 meV/Å (relative force error of 4.89\%), and a stress virial RMSE of 0.3 meV/Å$^3$ on test set.

\subsection{MD simulations}

Molecular dynamics (MD) simulations were performed using LAMMPS \cite{thompson2022lammps} with the MACE machine-learning interatomic potential \cite{batatia2022mace}. Periodic boundary conditions were applied in all three spatial directions, and a time step of 1 fs was used. The simulations were carried out in the isothermal--isobaric (NpT) ensemble using a Nos\'{e}--Hoover thermostat and barostat \cite{martyna1994constant}, with the temperature fixed at the target value and an external pressure of $-15$ kbar. The system was first equilibrated at 300 K for 150 ps and then cooled to the target temperature at a rate of 2.5 K/ps. A total of 180,000 MD steps, corresponding to 180 ps, were performed for each simulation after equilibration. Atomic configurations were recorded every 0.2 ps for subsequent analysis. OVITO \cite{stukowski2010visualization} was used to identify Pb-centered octahedra in the trajectories and calculate local polarization from the relative displacements of the Pb and Te centers and the corresponding Born effective charges. Correlation maps were interpolated onto a grid with four times finer spacing and Gaussian-smoothed in signed logarithmic space to extract the contours at a correlation threshold of 0.003.

Spectral energy density (SED) calculations were initialized using a fully equilibrated trajectory and the average unit cell dimensions from the production run. The system was first equilibrated at the target temperature in the NVT ensemble for 10 ps, followed by a microcanonical (NVE) production run of 131.072 ps with a time step of 1 fs. Atomic positions and velocities were recorded every 64 time steps. The SED was calculated using the dynasor package \cite{fransson2021dynasor}.

\subsection{IFC simulations}

The force-constant potential was constructed iteratively using structures generated by Monte Carlo rattling and phonon-based displacements, with the second-order force constants refined until the transverse optical (TO) phonon frequencies converged. Reference energies and forces were calculated using the trained MACE model, and the force constants were fitted using the hiphive package \cite{eriksson2019hiphive}. Molecular dynamics (MD) simulations were subsequently performed using the fitted IFC potentials in a 12 × 12 × 12 supercell, with third- and fourth-order terms included or excluded as specified in the main text. The simulations employed the NVT ensemble with a Langevin thermostat, a time step of $1~\mathrm{fs}$, and a friction coefficient of $0.02~\mathrm{fs}^{-1}$. The system was equilibrated at 200 K for 150 ps and cooled to the target temperature at a rate of 2.5 K/ps, after which a further $200~\mathrm{ps}$ production run was performed.

\begin{figure}
    \includegraphics[width=0.45\textwidth]{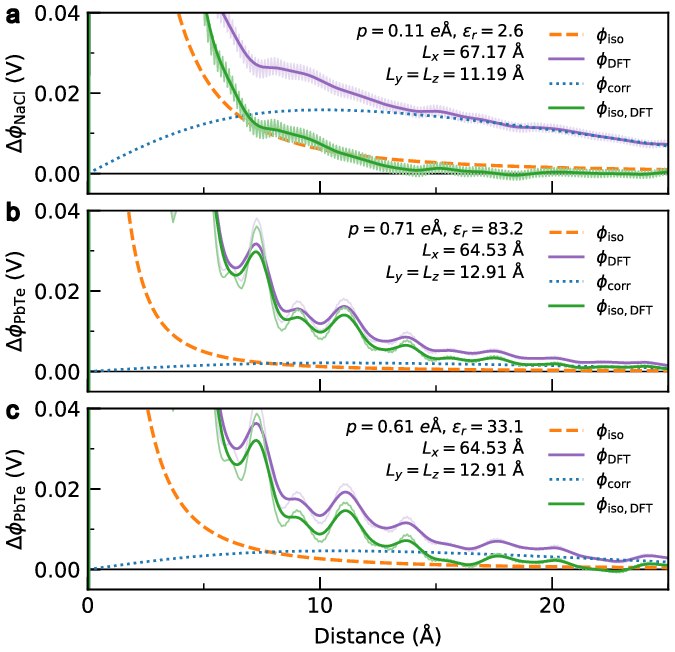}
        \caption{Correction of displacement-induced electrostatic potentials for periodic-image effects. The panels show (a) NaCl, (b) PbTe with spin--orbit coupling, and (c) PbTe without spin--orbit coupling. Dipole moments $p$, relative permittivities $\varepsilon_r$, and supercell dimensions $L_x$, $L_y$, and $L_z$ are indicated in each panel. Potentials are plotted as a function of distance along the displacement direction ($x$-axis).}
    \label{fig:potential_corr}
\end{figure}

\section{Dipole correction for periodic boundary conditions}

We calculate the difference between the electric potential of a periodic array of dipoles separated by $(L_x, L_y, L_z)$ and that of an isolated dipole. We start with the continuous charge density of a dipole derived from a normalized three-dimensional Gaussian function centered at the origin with variance $\sigma^2$, $g_\sigma(\mathbf r)$:
\begin{gather}
    g_\sigma(\mathbf r)=\frac{e^{-r^2/(2\sigma^2)}}{(2\pi\sigma^2)^{3/2}}.
\end{gather}
For a dipole oriented along $x$, the charge density is represented as
\begin{gather}
        \rho_{\rm{iso}}(\mathbf r)=-p\,\pdv{g_\sigma}{x}=\frac{px}{\sigma^2}g_\sigma(\mathbf r)
\end{gather}
where $p$ is the dipole moment.
The potential of an isolated dipole, $\phi_{\rm iso}$ can be represented using the error function as
\begin{gather}
    \phi_{\rm iso}(\mathbf r)=\frac{p}{4\pi\epsilon}\frac{x}{r^3}F(r),
\end{gather}
where
\begin{gather}
    F(r)=\operatorname{erf}\!\left(\frac{r}{\sqrt2\sigma}\right)
-\sqrt{\frac2\pi}\frac r\sigma e^{-r^2/(2\sigma^2)}.
\end{gather}
For $r \gg \sigma$, $\phi_{\rm iso}$ converges to
\begin{gather}
    \phi_{\rm iso}(\mathbf r)=\frac{p}{4\pi\epsilon}\frac{x}{r^3}.
\end{gather}

We now solve for the electric potential of repeated dipole images \cite{castro2003solution}, $\phi_{\rm repeat}$. Using the reciprocal lattice vector $\mathbf G=2\pi(n_x/L_x,n_y/L_y,n_z/L_z)$, the Fourier transform of the charge density $\rho_{\rm repeat}$ is
\begin{gather}
    \tilde\rho_{\rm repeat}(\mathbf G)=-ipG_xe^{-\sigma^2G^2/2}.
\end{gather}
Applying Poisson's equation $-\epsilon\nabla^2\phi_{\rm repeat}=\rho_{\rm repeat}$ in reciprocal space,
\begin{gather}
    \epsilon G^2\tilde \phi_{\rm repeat}(\mathbf G)=\tilde\rho_{\rm repeat}(\mathbf G) \\ 
    \phi_{\rm repeat}(\mathbf r)=\frac{p}{\epsilon V}\sum_{\mathbf G\ne0}\frac{-iG_x}{G^2}e^{-\sigma^2G^2/2}e^{i\mathbf G\cdot\mathbf r}
\end{gather}
where $V=L_xL_yL_z$ is the volume. The correction term $\phi_{\rm corr}$ to be subtracted from the DFT potential $\phi_{\rm DFT}$ to recover the isolated-dipole result $\phi_{\rm iso,DFT}$ is thus
\begin{gather}
    \phi_{\rm corr}(\mathbf r)=\frac{p}{\epsilon} \left (\frac{1}{V}\sum_{\mathbf G\ne0}\frac{-iG_x}{G^2}e^{-\sigma^2G^2/2}e^{i\mathbf G\cdot\mathbf r} -\frac{x}{4\pi r^3 }F(r) \right )
    \end{gather}
These three potentials, along with $\phi_{\rm iso}$, are shown in Fig.~\ref{fig:potential_corr} for NaCl and PbTe (with and without SOC).

\bibliography{references}

@article{bovzin2010entropically,
  title={Entropically stabilized local dipole formation in lead chalcogenides},
  author={Bo{\v{z}}in, Emil S and Malliakas, Christos D and Souvatzis, Petros and Proffen, Thomas and Spaldin, Nicola A and Kanatzidis, Mercouri G and Billinge, Simon J. L.},
  journal={Science},
  volume={330},
  number={6011},
  pages={1660--1663},
  year={2010},
  publisher={American Association for the Advancement of Science}
}

@article{sangiorgio2018correlated,
  title={Correlated local dipoles in {PbTe}},
  author={Sangiorgio, Boris and Bozin, Emil S and Malliakas, Christos D and Fechner, Michael and Simonov, Arkadiy and Kanatzidis, Mercouri G and Billinge, Simon J. L. and Spaldin, Nicola A and Weber, Thomas},
  journal={Physical Review Materials},
  volume={2},
  number={8},
  pages={085402},
  year={2018},
  publisher={APS}
}

@article{shiga2014origin,
  title={Origin of anomalous anharmonic lattice dynamics of lead telluride},
  author={Shiga, Takuma and Murakami, Takuru and Hori, Takuma and Delaire, Olivier and Shiomi, Junichiro},
  journal={Applied Physics Express},
  volume={7},
  number={4},
  pages={041801},
  year={2014},
  publisher={The Japan Society of Applied Physics}
}

@article{jensen2012lattice,
  title={Lattice dynamics reveals a local symmetry breaking in the emergent dipole phase of {PbTe}},
  author={Jensen, Kirsten M. {\O}. and Bo{\v{z}}in, Emil S and Malliakas, Christos D and Stone, Matthew B and Lumsden, Mark D and Kanatzidis, Mercouri G and Shapiro, Stephen M and Billinge, Simon J. L.},
  journal={Physical Review B},
  volume={86},
  number={8},
  pages={085313},
  year={2012},
  publisher={APS}
}

@article{chen2014first,
  title={{First-principles approach to nonlinear lattice dynamics: Anomalous spectra in PbTe}},
  author={Chen, Yue and Ai, Xinyuan and Marianetti, C. A.},
  journal={Physical Review Letters},
  volume={113},
  number={10},
  pages={105501},
  year={2014},
  publisher={APS}
}

@article{li2022anomalous,
  title={{Anomalous transverse optical phonons in SnTe and PbTe}},
  author={Li, Zehua and Li, Shasha and Castellan, John-Paul and Heid, Rolf and Xiao, Yu and Zhao, Li-Dong and Chen, Yue and Weber, Frank},
  journal={Physical Review B},
  volume={105},
  number={1},
  pages={014308},
  year={2022},
  publisher={APS}
}

@article{tacke1982dielectric,
  title={{The dielectric constant of PbTe at 4.2 K and $\tilde{\nu}= 84.15$ cm$^{-1}$, 96.97 cm$^{-1}$, 103.60 cm$^{-1}$}},
  author={Tacke, M and Schuberth, W and Becker, Charles R and Haas, L. D.},
  journal={Applied Physics A},
  volume={28},
  pages={229--233},
  year={1982}
}

@inproceedings{schutt2017schnet,
  title={{SchNet}: A continuous-filter convolutional neural network for modeling quantum interactions},
  author={Sch{\"u}tt, Kristof and Kindermans, Pieter-Jan and Sauceda Felix, Huziel Enoc and Chmiela, Stefan and Tkatchenko, Alexandre and M{\"u}ller, Klaus-Robert},
  booktitle={Advances in Neural Information Processing Systems},
  volume={30},
  year={2017}
}

@inproceedings{gasteiger2020directional,
  title={Directional Message Passing for Molecular Graphs},
  author={Johannes Gasteiger and Janek Gro{\ss} and Stephan G{\"u}nnemann},
  booktitle={International Conference on Learning Representations},
  year={2020},
  NOurl={https://openreview.net/forum?id=B1eWbxStPH}
}

@inproceedings{batatia2022mace,
  title={{MACE}: Higher Order Equivariant Message Passing Neural Networks for Fast and Accurate Force Fields},
  author={Batatia, Ilyes and Kov{\'a}cs, D{\'a}vid P{\'e}ter and Simm, Gregor N. C. and Ortner, Christoph and Cs{\'a}nyi, G{\'a}bor},
  booktitle={Advances in Neural Information Processing Systems},
  volume={35},
  year={2022},
  NOurl={https://openreview.net/forum?id=YPpSngE-ZU}
}

@article{Batatia2022Design,
  title = {The Design Space of {E(3)}-Equivariant Atom-Centered Interatomic Potentials},
  author = {Batatia, Ilyes and Batzner, Simon and Kov{\'a}cs, D{\'a}vid P{\'e}ter and Musaelian, Albert and Simm, Gregor N. C. and Drautz, Ralf and Ortner, Christoph and Kozinsky, Boris and Cs{\'a}nyi, G{\'a}bor},
  journal={Nature Machine Intelligence},
  volume={7},
  number={1},
  pages={56--67},
  year={2025},
  publisher={Nature Publishing Group UK, London}
 }

@article{eriksson2019hiphive,
  title={The {hiphive} package for the extraction of high-order force constants by machine learning},
  author={Eriksson, Fredrik and Fransson, Erik and Erhart, Paul},
  journal={Advanced Theory and Simulations},
  volume={2},
  number={5},
  pages={1800184},
  year={2019},
  publisher={Wiley Online Library}
}

@article{kresse1996efficient,
  title={Efficient iterative schemes for ab initio total-energy calculations using a plane-wave basis set},
  author={Kresse, Georg and Furthm{\"u}ller, J{\"u}rgen},
  journal={Physical Review B},
  volume={54},
  number={16},
  pages={11169},
  year={1996},
  publisher={APS}
}

@article{kresse1999ultrasoft,
  title={From ultrasoft pseudopotentials to the projector augmented-wave method},
  author={Kresse, Georg and Joubert, Daniel},
  journal={Physical Review B},
  volume={59},
  number={3},
  pages={1758},
  year={1999},
  publisher={APS}
}

@article{perdew1981self,
  title={Self-interaction correction to density-functional approximations for many-electron systems},
  author={Perdew, John P and Zunger, Alex},
  journal={Physical Review B},
  volume={23},
  number={10},
  pages={5048},
  year={1981},
  publisher={APS}
}

@article{blochl1994projector,
  title={Projector augmented-wave method},
  author={Bl{\"o}chl, Peter E},
  journal={Physical Review B},
  volume={50},
  number={24},
  pages={17953},
  year={1994},
  publisher={APS}
}

@article{jinnouchi2019phase,
  title={Phase transitions of hybrid perovskites simulated by machine-learning force fields trained on the fly with {Bayesian inference}},
  author={Jinnouchi, Ryosuke and Lahnsteiner, Jonathan and Karsai, Ferenc and Kresse, Georg and Bokdam, Menno},
  journal={Physical Review Letters},
  volume={122},
  number={22},
  pages={225701},
  year={2019},
  publisher={APS}
}

@article{monkhorst1976special,
  title={{Special points for Brillouin-zone integrations}},
  author={Monkhorst, Hendrik J and Pack, James D},
  journal={Physical Review B},
  volume={13},
  number={12},
  pages={5188},
  year={1976},
  publisher={APS}
}

@article{perdew2008restoring,
  title={Restoring the density-gradient expansion for exchange in solids and surfaces},
  author={Perdew, John P and Ruzsinszky, Adrienn and Csonka, G{\'a}bor I and Vydrov, Oleg A and Scuseria, Gustavo E and Constantin, Lucian A and Zhou, Xiaolan and Burke, Kieron},
  journal={Physical Review Letters},
  volume={100},
  number={13},
  pages={136406},
  year={2008},
  publisher={APS}
}

@article{minikayev2021thermostructural,
  title={{Thermostructural and elastic properties of PbTe and Pb$_{0.884}$Cd$_{0.116}$Te: A combined low-temperature and high-pressure x-ray diffraction study of Cd-substitution effects}},
  author={Minikayev, Roman and Safari, Fatemeh and Katrusiak, Andrzej and Szuszkiewicz, Wojciech and Szczerbakow, Andrzej and Bell, Anthony and Dynowska, El{\.z}bieta and Paszkowicz, Wojciech},
  journal={Crystals},
  volume={11},
  number={9},
  pages={1063},
  year={2021},
  publisher={MDPI}
}

@article{van1999correcting,
  title={Correcting overbinding in local-density-approximation calculations},
  author={Van de Walle, A and Ceder, G},
  journal={Physical Review B},
  volume={59},
  number={23},
  pages={14992},
  year={1999},
  publisher={APS}
}

@article{cochran1966crystal,
  title={The crystal dynamics of lead telluride},
  author={Cochran, William and Cowley, Roger Arthur and Dolling, G and Elcombe, M. M.},
  journal={Proceedings of the Royal Society of London. Series A, Mathematical and Physical Sciences},
  volume={293},
  pages={433--451},
  year={1966},
  publisher={JSTOR}
}

@article{ravindra1980properties,
  title={{Properties of PbS, PbSe, and PbTe}},
  author={Ravindra, N. M. and Srivastava, V. K.},
  journal={physica status solidi (a)},
  volume={58},
  number={1},
  pages={311--316},
  year={1980},
  publisher={Wiley-VCH Verlag}
}

@article{romero2015thermal,
  title={{Thermal conductivity in PbTe from first principles}},
  author={Romero, A. H. and Gross, E. K. U. and Verstraete, M. J. and Hellman, Olle},
  journal={Physical Review B},
  volume={91},
  number={21},
  pages={214310},
  year={2015},
  publisher={APS}
}

@article{skelton2014thermal,
  title={{Thermal physics of the lead chalcogenides PbS, PbSe, and PbTe from first principles}},
  author={Skelton, Jonathan M and Parker, Stephen C and Togo, Atsushi and Tanaka, Isao and Walsh, Aron},
  journal={Physical Review B},
  volume={89},
  number={20},
  pages={205203},
  year={2014},
  publisher={APS}
}

@article{park2022off,
  title={{Off-centered Pb interstitials in PbTe}},
  author={Park, Sungjin and Ryu, Byungki and Park, SuDong},
  journal={Materials},
  volume={15},
  number={4},
  pages={1272},
  year={2022},
  publisher={MDPI}
}

@article{song2017ab,
  title={Ab initio study of electron mean free paths and thermoelectric properties of lead telluride},
  author={Song, Qichen and Liu, Te-Huan and Zhou, Jiawei and Ding, Zhiwei and Chen, Gang},
  journal={Materials Today Physics},
  volume={2},
  pages={69--77},
  year={2017},
  publisher={Elsevier}
}

@article{an2008ab,
  title={{Ab initio phonon dispersions for PbTe}},
  author={An, Jiming and Subedi, Alaska and Singh, David J},
  journal={Solid State Communications},
  volume={148},
  number={9--10},
  pages={417--419},
  year={2008},
  publisher={Elsevier}
}

@article{delaire2011giant,
  title={{Giant anharmonic phonon scattering in PbTe}},
  author={Delaire, Olivier and Ma, Jie and Marty, Karol and May, Andrew F and McGuire, Michael A and Du, Mao-Hua and Singh, David J and Podlesnyak, A and Ehlers, G and Lumsden, M. D. and others},
  journal={Nature Materials},
  volume={10},
  number={8},
  pages={614--619},
  year={2011},
  publisher={Nature Publishing Group UK London}
}

@article{tian2012phonon,
  title={{Phonon conduction in PbSe, PbTe, and PbTe$_{1-x}$Se$_x$ from first-principles calculations}},
  author={Tian, Zhiting and Garg, Jivtesh and Esfarjani, Keivan and Shiga, Takuma and Shiomi, Junichiro and Chen, Gang},
  journal={Physical Review B},
  volume={85},
  number={18},
  pages={184303},
  year={2012},
  publisher={APS}
}

@article{querales2019temperature,
  title={{Temperature effects on the electronic band structure of PbTe from first principles}},
  author={Querales-Flores, Jos{\'e} D and Cao, Jiang and Fahy, Stephen and Savi{\'c}, Ivana},
  journal={Physical Review Materials},
  volume={3},
  number={5},
  pages={055405},
  year={2019},
  publisher={APS}
}

@article{gibbs2013temperature,
  title={{Temperature dependent band gap in PbX (X=S, Se, Te)}},
  author={Gibbs, Zachary M and Kim, Hyoungchul and Wang, Heng and White, Robert L and Drymiotis, Fivos and Kaviany, Massoud and Snyder, G. Jeffrey},
  journal={Applied Physics Letters},
  volume={103},
  number={26},
  year={2013},
  pages={262109},
  publisher={AIP Publishing}
}

@article{thompson2022lammps,
  title={{LAMMPS---a flexible simulation tool for particle-based materials modeling at the atomic, meso, and continuum scales}},
  author={Thompson, Aidan P and Aktulga, H Metin and Berger, Richard and Bolintineanu, Dan S and Brown, W Michael and Crozier, Paul S and In't Veld, Pieter J and Kohlmeyer, Axel and Moore, Stan G and Nguyen, Trung Dac and others},
  journal={Computer Physics Communications},
  volume={271},
  pages={108171},
  year={2022},
  publisher={Elsevier}
}

@article{martyna1994constant,
  title={Constant pressure molecular dynamics algorithms},
  author={Martyna, Glenn J and Tobias, Douglas J and Klein, Michael L},
  journal={The Journal of Chemical Physics},
  volume={101},
  pages={4177},
  year={1994}
}

@article{halasyamani1998noncentrosymmetric,
  title={Noncentrosymmetric oxides},
  author={Halasyamani, P Shiv and Poeppelmeier, Kenneth R},
  journal={Chemistry of Materials},
  volume={10},
  number={10},
  pages={2753--2769},
  year={1998},
  publisher={ACS Publications}
}

@article{kusainova2001ferroelectric,
  title={Ferroelectric properties and crystal structure of the layered intergrowth phase {Bi$_3$Pb$_2$Nb$_2$O$_{11}$Cl}},
  author={Kusainova, Ardak M and Lightfoot, Philip and Zhou, Wuzong and Stefanovich, Sergei Yu and Mosunov, Alexander V and Dolgikh, Valerii A},
  journal={Chemistry of Materials},
  volume={13},
  number={12},
  pages={4731--4737},
  year={2001},
  publisher={ACS Publications}
}

@article{niesner2018structural,
  title={Structural fluctuations cause spin-split states in tetragonal {(CH$_3$NH$_3$)PbI$_3$} as evidenced by the circular photogalvanic effect},
  author={Niesner, Daniel and Hauck, Martin and Shrestha, Shreetu and Levchuk, Ievgen and Matt, Gebhard J and Osvet, Andres and Batentschuk, Miroslaw and Brabec, Christoph and Weber, Heiko B and Fauster, Thomas},
  journal={Proceedings of the National Academy of Sciences},
  volume={115},
  number={38},
  pages={9509--9514},
  year={2018},
  publisher={National Academy of Sciences}
}

@article{zhang2026imaging,
  title={Imaging of nanoscale polar textures in quantum paraelectric {SrTiO$_3$}},
  author={Zhang, Yang and Sung, Suk Hyun and Agarwal, Nishkarsh and Gates, Maya and Li, Cong and Yu, Pu and Hovden, Robert and El Baggari, Ismail},
  journal={Nature},
  volume={656},
  pages={54--59},
  year={2026},
  publisher={Nature Publishing Group UK London}
}

@article{orenstein2025observation,
  title={Observation of polarization density waves in {SrTiO$_3$}},
  author={Orenstein, Gal and Krapivin, Viktor and Huang, Yijing and Zhang, Zhuquan and de la Pe{\~n}a Mu{\~n}oz, Gilberto and Duncan, Ryan A and Nguyen, Quynh and Stanton, Jade and Teitelbaum, Samuel and Yavas, Hasan and others},
  journal={Nature Physics},
  volume={21},
  number={6},
  pages={961--965},
  year={2025},
  publisher={Nature Publishing Group UK London}
}

@article{gonze1997dynamical,
  title={{Dynamical matrices, Born effective charges, dielectric permittivity tensors, and interatomic force constants from density-functional perturbation theory}},
  author={Gonze, Xavier and Lee, Changyol},
  journal={Physical Review B},
  volume={55},
  number={16},
  pages={10355},
  year={1997},
  publisher={APS}
}

@article{knox2014local,
  title={Local off-centering symmetry breaking in the high-temperature regime of {SnTe}},
  author={Knox, K. R. and Bozin, E. S. and Malliakas, C. D. and Kanatzidis, M. G. and Billinge, S. J. L.},
  journal={Physical Review B},
  volume={89},
  number={1},
  pages={014102},
  year={2014}
}

@article{bridges2023local,
  title={{Local structure, bonding, and asymmetry of ((NH$_2$)$_2$CH)PbBr$_3$, CsPbBr$_3$, and (CH$_3$NH$_3$)PbBr$_3$}},
  author={Bridges, F and Gruzdas, J and MacKeen, C and Mayford, K and Weadock, N. J. and {Urena Baltazar}, V. and Rakita, Y and Waquier, Louis and Vigil, Julian A and Karunadasa, Hemamala I and Toney, M. F.},
  journal={Physical Review B},
  volume={108},
  number={21},
  pages={214102},
  year={2023}
}

@article{cochran1959crystal,
  title={Crystal stability and the theory of ferroelectricity},
  author={Cochran, W.},
  journal={Physical Review Letters},
  volume={3},
  number={9},
  pages={412},
  year={1959},
  publisher={APS}
}

@inproceedings{anderson1960qualitative,
  author       = {Anderson, P. W.},
  title        = {Qualitative Considerations on the Statistics of the Phase Transition in {BaTiO$_3$}-type Ferroelectrics},
  booktitle    = {Fizika Dielektrikov, Conference Proceedings, Lebedev Physics Institute (Nov. 1958)},
  pages        = {290},
  year         = {1960},
  publisher    = {Academy of Sciences of the USSR},
  address      = {Moscow}
}

@article{fauque2022mesoscopic,
  title={Mesoscopic fluctuating domains in strontium titanate},
  author={Fauqu{\'e}, Beno{\^\i}t and Bourges, Philippe and Subedi, Alaska and Behnia, Kamran and Baptiste, Beno{\^\i}t and Roessli, Bertrand and Fennell, Tom and Raymond, St{\'e}phane and Steffens, Paul},
  journal={Physical Review B},
  volume={106},
  number={14},
  pages={L140301},
  year={2022},
  publisher={APS}
}

@book{born1954dynamical,
  title={Dynamical theory of crystal lattices},
  author={Born, Max and Huang, Kun},
  year={1954},
  publisher={Oxford University Press}
}

@article{fransson2021dynasor,
  title={{dynasor}---a tool for extracting dynamical structure factors and current correlation functions from molecular dynamics simulations},
  author={Fransson, Erik and Slabanja, Mattias and Erhart, Paul and Wahnstr{\"o}m, G{\"o}ran},
  journal={Advanced Theory and Simulations},
  volume={4},
  number={2},
  pages={2000240},
  year={2021},
  publisher={Wiley Online Library}
}

@article{stukowski2010visualization,
  title={Visualization and analysis of atomistic simulation data with {OVITO}---the {Open Visualization Tool}},
  author={Stukowski, Alexander},
  journal={Modelling and Simulation in Materials Science and Engineering},
  volume={18},
  number={1},
  pages={015012},
  year={2010}
}

@article{keiber2013lead,
  title={{Lead is not off center in PbTe: the importance of $r$-space phase information in extended X-ray absorption fine structure spectroscopy}},
  author={Keiber, T and Bridges, F and Sales, B. C.},
  journal={Physical Review Letters},
  volume={111},
  number={9},
  pages={095504},
  year={2013},
  publisher={APS}
}

@article{manley2016giant,
  title={Giant electromechanical coupling of relaxor ferroelectrics controlled by polar nanoregion vibrations},
  author={Manley, Michael E and Abernathy, Douglas L and Sahul, Raffi and Parshall, Daniel E and Lynn, Jeffrey W and Christianson, Andrew D and Stonaha, Paul J and Specht, Eliot D and Budai, John D},
  journal={Science Advances},
  volume={2},
  number={9},
  pages={e1501814},
  year={2016},
  publisher={American Association for the Advancement of Science}
}

@article{riemelmoser2026dielectric,
  title={Dielectric-dependent hybrid functional based on meta-{GGA}},
  author={Riemelmoser, Stefan and Xu, Xun and Pasquarello, Alfredo},
  journal={Nature Communications},
  year={2026},
  volume={17},
  pages={8323},
  publisher={Nature Publishing Group UK, London}
}

@article{ju2018revisiting,
  title={Revisiting {PbTe} to identify how thermal conductivity is really limited},
  author={Ju, Shenghong and Shiga, Takuma and Feng, Lei and Shiomi, Junichiro},
  journal={Physical Review B},
  volume={97},
  number={18},
  pages={184305},
  year={2018},
  publisher={APS}
}

@article{lee2014resonant,
  title={Resonant bonding leads to low lattice thermal conductivity},
  author={Lee, Sangyeop and Esfarjani, Keivan and Luo, Tengfei and Zhou, Jiawei and Tian, Zhiting and Chen, Gang},
  journal={Nature Communications},
  volume={5},
  number={1},
  pages={3525},
  year={2014},
  publisher={Nature Publishing Group UK, London}
}

@article{wuttig2018incipient,
  title={Incipient metals: functional materials with a unique bonding mechanism},
  author={Wuttig, Matthias and Deringer, Volker L and Gonze, Xavier and Bichara, Christophe and Raty, Jean-Yves},
  journal={Advanced Materials},
  volume={30},
  number={51},
  pages={1803777},
  year={2018},
  publisher={Wiley Online Library}
}

@article{hempelmann2021long,
  title={Long-Range Forces in Rock-Salt-Type Tellurides and How they Mirror the Underlying Chemical Bonding},
  author={Hempelmann, Jan and M{\"u}ller, Peter C and Konze, Philipp M and Stoffel, Ralf P and Steinberg, Simon and Dronskowski, Richard},
  journal={Advanced Materials},
  volume={33},
  number={37},
  pages={2100163},
  year={2021},
  publisher={Wiley Online Library}
}

@article{rubtsov2000crossover,
  title={Crossover between a displacive and an order-disorder phase transition},
  author={Rubtsov, A. N. and Hlinka, J and Janssen, T},
  journal={Physical Review E},
  volume={61},
  number={1},
  pages={126},
  year={2000},
  publisher={APS}
}

@article{blinc2004order,
  title={Order and disorder in ferroelectrics},
  author={Blinc, R},
  journal={Ferroelectrics},
  volume={301},
  number={1},
  pages={3--8},
  year={2004},
  publisher={Taylor \& Francis}
}

@article{guzman2023lamellar,
  title={Lamellar fluctuations melt ferroelectricity},
  author={Guzm{\'a}n-Verri, Gian Giacomo and Liang, Charles H and Littlewood, Peter B},
  journal={Physical Review Letters},
  volume={131},
  number={4},
  pages={046801},
  year={2023},
  publisher={APS}
}

@article{marais1991phenomena,
  title={Phenomena due to strain coupling in phase transitions},
  author={Marais, Simon and Heine, Volker and Nex, Chris and Salje, Ekhard},
  journal={Physical Review Letters},
  volume={66},
  number={19},
  pages={2480},
  year={1991},
  publisher={APS}
}

@article{kawahala2023thickness,
  title={Thickness-dependent terahertz permittivity of epitaxially grown {PbTe} thin films},
  author={Kawahala, Nicolas M and Matos, Daniel A and Rappl, Paulo H. O. and Abramof, Eduardo and Baydin, Andrey and Kono, Junichiro and Hernandez, Felix G. G.},
  journal={Coatings},
  volume={13},
  number={11},
  pages={1855},
  year={2023},
  publisher={MDPI}
}

@article{strehlow1973compilation,
  title={Compilation of energy band gaps in elemental and binary compound semiconductors and insulators},
  author={Strehlow, W. H. and Cook, Earl L},
  journal={Journal of Physical and Chemical Reference Data},
  volume={2},
  number={1},
  pages={163--200},
  year={1973},
  publisher={American Institute of Physics}
}

@article{castro2003solution,
  title={Solution of {Poisson}'s equation for finite systems using plane-wave methods},
  author={Castro, Alberto and Rubio, Angel and Stott, M. J.},
  journal={Canadian Journal of Physics},
  volume={81},
  number={10},
  pages={1151--1164},
  year={2003},
  publisher={NRC Research Press, Ottawa, Canada}
}

@inproceedings{kosmala2023ewald,
  title={Ewald-based long-range message passing for molecular graphs},
  author={Kosmala, Arthur and Gasteiger, Johannes and Gao, Nicholas and G{\"u}nnemann, Stephan},
  booktitle={International Conference on Machine Learning},
  pages={17544--17563},
  year={2023},
  organization={PMLR}
}

@article{waghmare2003first,
  title={{First-principles indicators of metallicity and cation off-centricity in the IV-VI rocksalt chalcogenides of divalent Ge, Sn, and Pb}},
  author={Waghmare, UV and Spaldin, NA and Kandpal, HemChandra and Seshadri, Ram},
  journal={Physical Review B},
  volume={67},
  number={12},
  pages={125111},
  year={2003},
  publisher={APS}
}

@article{bersuker2013pseudo,
  title={{Pseudo-Jahn--Teller} Effect---A Two-State Paradigm in Formation, Deformation, and Transformation of Molecular Systems and Solids},
  author={Bersuker, Isaac B},
  journal={Chemical Reviews},
  volume={113},
  number={3},
  pages={1351--1390},
  year={2013},
  publisher={ACS Publications}
}

@article{onodera1970dynamic,
  title={Dynamic susceptibility of classical anharmonic oscillator: a unified oscillator model for order-disorder and displacive ferroelectrics},
  author={Onodera, Yositaka},
  journal={Progress of Theoretical Physics},
  volume={44},
  number={6},
  pages={1477--1499},
  year={1970},
  publisher={Oxford University Press}
}

\end{document}


\title{
\textbf{Harmonic origin of local symmetry breaking in PbTe revealed using machine-learned potentials\\[2ex] \normalsize \textsl{Supplementary Information}}
}

\author{Seongjoo Jung}
\author{Nicola A. Spaldin}

\affiliation{Department of Materials, ETH Z\"{u}rich, Wolfgang-Pauli-Strasse 27, 8093 Z\"{u}rich, Switzerland
}

\maketitle

\section{Choice of DFT functional}

PbTe requires careful consideration in DFT simulations due to its combination of low optical phonon frequencies, a small bandgap, and the resulting slow convergence. 
Previous studies have employed a range of exchange-correlation functionals, including AM05 \cite{romero2015thermal}, PBEsol \cite{skelton2014thermal}, PBE \cite{park2022off,li2022anomalous}, LDA \cite{an2008ab,bovzin2010entropically,tian2012phonon,song2017ab}, r$^2$SCAN \cite{riemelmoser2026dielectric}, TPSS \cite{skelton2014thermal} and HSE03 \cite{skelton2014thermal}
. We therefore perform a assessment of selective levels of exchange-correlation functionals, ranging from the local density approximation (LDA) to meta-GGA functionals, both with and without spin-orbit coupling (SOC). 

\begin{figure*}[!htb]
\includegraphics[width=0.90\textwidth]{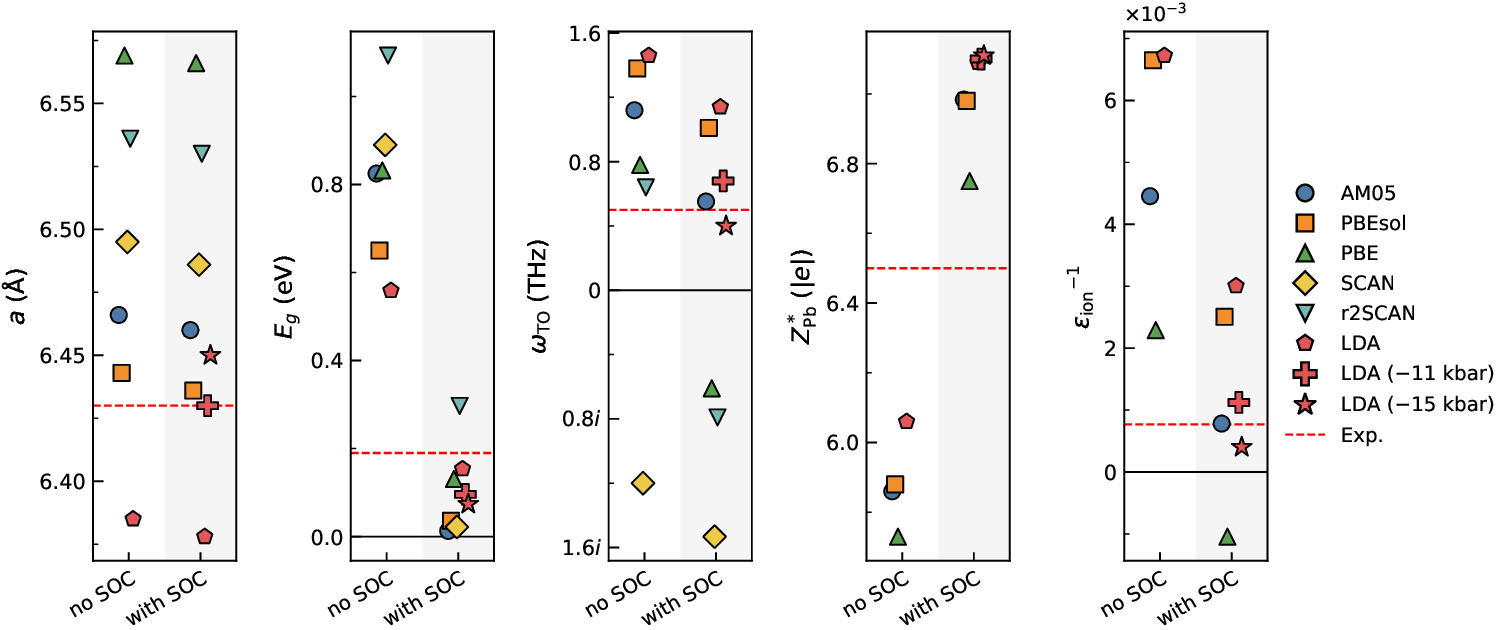}
\caption{Comparison of lattice constant ($a$), bandgap ($E_g$), $\Gamma$-point transverse optical phonon frequency ($\omega_{\text{TO}}$), Born effective charge of lead ($Z^*_{\text{Pb}}$), and the inverse ionic dielectric constant ($\varepsilon_{\text{ion}}^{-1}$) of PbTe calculated using various density functional theory exchange-correlation functionals, both without and with the inclusion of spin-orbit coupling. Red dashed lines represent target experimental values.}
\label{fig:xc}
\end{figure*}

Unfortunately, as shown in Fig.~\ref{fig:xc}, none of the exchange-correlation functionals considered provides a consistently accurate description of all target properties. In general, the inclusion of SOC produces noticeable changes in the calculated properties and, in most cases, brings the results into closer agreement with experiment. Among the GGA functionals, AM05 with SOC provides a reasonable description of the lattice constant, TO phonon frequencies, and dielectric response. However, the band gap is nearly closed, placing the system too close to a metallic transition. The meta-GGA functional r$^2$SCAN with SOC yields band gaps and dielectric response in good agreement with experiment, but produces imaginary TO phonon frequencies, indicating a dynamical instability. LDA with SOC provides a satisfactory description of phonon band structures with a robust band gap, although it exhibits the characteristic overbinding associated with the LDA as well as reported band symmetry inversion localized at L point \cite{querales2019temperature}.

The overall result indicates that accurate description of phononic and electronic band structure in PbTe requires separate exchange-correlation treatment, or use of hybrid functionals \cite{skelton2014thermal}. For the purpose of this study, we prioritize accurate description of the molecular dynamics properties associated with the TO mode and computationally viable methods to create enough dataset for MLIP trainnig. Ultimately, we adopt the LDA functional with SOC and apply a negative pressure to compensate for its intrinsic overbinding tendency \cite{van1999correcting} which provides a reasonable overall balance among the structural, electronic, and phononic properties of PbTe.

\section{DFT Convergence}

\begin{figure}[!htb]
\includegraphics[width=0.40\textwidth]{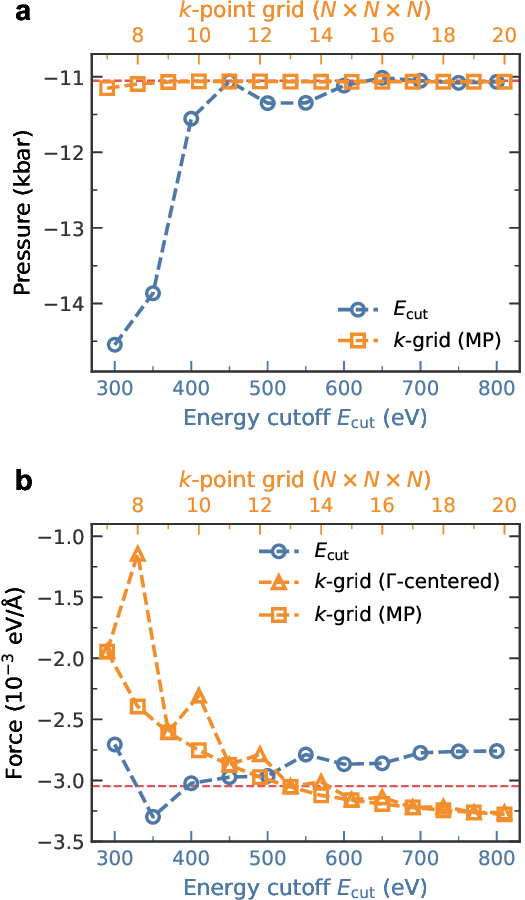}
\caption{Parametric convergence of (a) pressure at cubic phase and (b) forces on Pb corresponding to TO mode displacement of 0.02 Å in PbTe. Red dahsed line indicate value at full convergence (700 eV energy cutoff and $18\times18\times18$ MP $k$-grid)
} \label{fig:convergence}
\end{figure}

We performed pressure convergence tests using a cubic unit cell with $a=6.430$~Å, and force convergence tests using a collective Pb (TO mode) displacement of 0.02~Å for the conventional unit cell of PbTe, within the LDA+$\mathrm{SOC}$ framework. The results are shown in Fig.~\ref{fig:convergence}. For the $k$-point convergence test, $E_\mathrm{cut}$ was fixed at 450~eV, while a $12\times12\times12$ $k$-point grid was used for the $E_\mathrm{cut}$ convergence test. Panel~(a) shows that the calculated pressure is relatively insensitive to the $k$-point density but highly sensitive to $E_\mathrm{cut}$. Although the pressure fluctuates around its converged value near $E_\mathrm{cut}$ of 450 eV, full convergence requires a substantially higher cutoff of approximately 600 eV.

For force convergence, shown in panel~(b), both a high $E_\mathrm{cut}$ and a dense $k$-point grid are required. Note that even-numbered $\Gamma$-centered $k$-point grids which include high-symmetry points at the Brillouin-zone boundary exhibit notably slower convergence than the Monkhorst--Pack (MP) grid. Full convergence is achieved near $E_\mathrm{cut}$ of 700 eV and a $18\times18\times18$ MP grid, which is computationally not feasible for generating enough data for MLIP training. Since lower values of $E_\mathrm{cut}$ systematically overestimate the forces, while coarser $k$-point grids systematically underestimate them, we adopted $E_\mathrm{cut}$ of 450 eV and a $12\times12\times12$ MP grid as a compromise between computational cost and accuracy, yielding values that are sufficiently close to the fully converged results indicated by red horizontal lines. 

\newpage

\section{third-NN cutoff calculations}

\begin{figure*}[!htb]
\includegraphics[width=1\textwidth]{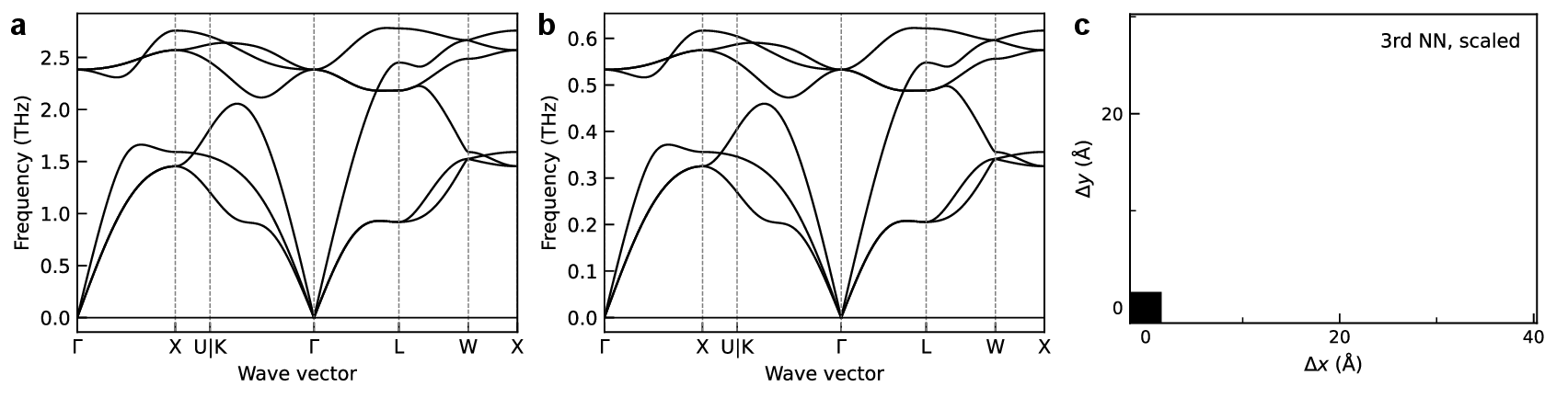}
\caption{(a) Harmonic phonon dispersion relations constructed using interaction cutoffs at the third nearest neighbors. (b) Phonon dispersion resulting from scaling down the harmonic interactions by a factor of 0.05. (c) Polar correlation from scaled down IFCs, where $f_{\rm TO}=0.53$ THz.}
\label{fig:3rdNN}
\end{figure*}

Fig.~\ref{fig:3rdNN} (a) shows the phonon dispersion where IFCs up to the third nearest-neighbor interaction are included, preserving the acoustic sum rule. The TO mode is noticeably harder at $f_{\rm TO}=2.4$ THz, and the TO band is flatter. Scaling the IFCs uniformly down by a factor of 0.05 results in phonon dispersion plot, where frequency of every mode is also uniformly scaled down, as shown in panel (b). Despite having a TO mode frequency corresponding to long-range interaction cases at $f_{\rm TO}=0.53$ THz, the polar correlations completely vanish even for the nearest octahedron using only harmonic force constants at 200 K.

\FloatBarrier
\bibliography{references}